\documentclass[%
reprint,
superscriptaddress,
sort&compress,
 amsmath,amssymb,
prd,
]{revtex4-2}
\usepackage{enumitem}
\usepackage[utf8]{inputenc}
\usepackage{comment}
\usepackage{xcolor}
\usepackage{hyperref}
\usepackage{graphicx}
\usepackage{dcolumn}
\usepackage{bm}
\usepackage{amsmath}
\usepackage{makecell}
\DeclareMathOperator*{\argmax}{arg\,max}

\graphicspath{{figures/}{./}}

\def\prd{{\it Phys. Rev.} D~}

\def\apjl{{\it Astrophys. J.} Lett~}
\def\apj{{\it Astrophys. J.}}

\def\mnras{{\it MNRAS~}}

\def\apj{{\it Astrophys. J.~}}
\def\apjl{{\it Astrophys. J. Lett.~}}
\def\nat{{\it Nature~}}

\newcommand{\orcidlink}[1]{\href{https://orcid.org/#1}{\includegraphics[width=10pt]{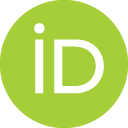}}}

\begin{document}

\title{Numerical relativity higher order gravitational waveforms of eccentric, spinning, nonprecessing binary black hole mergers}

\author{Abhishek V. Joshi \orcidlink{0000-0002-2514-5965}}
\email{avjoshi2@illinois.edu}
\affiliation{ 
NCSA, University of Illinois at Urbana-Champaign, Urbana, Illinois 61801, USA
}%
\affiliation{ 
Department of Physics, University of Illinois at Urbana-Champaign, Urbana, Illinois 61801, USA
}%
\author{Shawn G. Rosofsky}
\affiliation{ 
NCSA, University of Illinois at Urbana-Champaign, Urbana, Illinois 61801, USA
}%
\affiliation{ 
Department of Physics, University of Illinois at Urbana-Champaign, Urbana, Illinois 61801, USA
}%

\author{Roland Haas \orcidlink{0000-0003-1424-6178}}
\affiliation{ 
NCSA, University of Illinois at Urbana-Champaign, Urbana, Illinois 61801, USA
}%
\affiliation{ 
Department of Physics, University of Illinois at Urbana-Champaign, Urbana, Illinois 61801, USA
}%

\author{E. A. Huerta \orcidlink{0000-0002-9682-3604}}
\affiliation{ 
Data Science and Learning Division, Argonne National Laboratory, Lemont, Illinois 60439,
USA
}%
\affiliation{ 
Department of Computer Science, University of Chicago, Chicago, Illinois 60637, USA
}%
\affiliation{ 
Department of Physics, University of Illinois at Urbana-Champaign, Urbana, Illinois 61801, USA
}%

\begin{abstract}
\noindent We use the open source, community-driven, numerical 
relativity software, the Einstein Toolkit to study the physics 
of eccentric, spinning, nonprecessing binary black hole mergers 
with mass-ratios \(q=\{2, 4, 6\}\), individual dimensionless
spin parameters \(\chi_{1z}=\pm0.6\), \(\chi_{2z}=\pm0.3\), that include higher order gravitational wave modes \(\ell\leq4\), 
except for memory modes. 
Assuming stellar mass binary black hole mergers that may be 
detectable by the advanced LIGO detectors, we find that 
including modes up to \(\ell=4\) increases the signal-to-noise
of compact binaries between \(3.5\%\) to \(35\%\), compared to 
signals that only include the \(\ell=|m|=2\) mode. We 
use two waveform models, 
TEOBResumS and SEOBNRE, which incorporate 
spin and eccentricity corrections in the waveform dynamics, 
to quantify the orbital eccentricity of our 
numerical relativity catalog in a gauge-invariant manner 
through fitting factor calculations. Our findings indicate 
that the inclusion 
of higher order wave modes has a measurable effect in the 
recovery of moderately and highly eccentric black hole mergers, 
and thus it is essential to develop waveform models and 
signal processing tools that accurately describe the 
physics of these astrophysical sources.
\end{abstract}


\maketitle

\section{Introduction}

\noindent The modeling of eccentric 
compact binary mergers  
has attracted significant attention in recent years. 
The understanding of these astrophysical sources has 
gradually increased through a variety of 
analytical and numerical relativity studies 
that have shed new light into physics of 
these systems, and the properties of the 
gravitational wave signals that may be emitted 
by these 
sources~\cite{moore:2018fde,ramos_buades,huerta_nr_catalog,hinder:2017a,cao:2017,Hinderer:2017,ENIGMA_Huerta,2021PhRvD.103h4018C,Huerta:2017a,Osburn:2016,lou:2016arXiv,lou:2017CQG,ihh:2008PhRvD,binithi:2016PhRvD,kavbini:2017PhRvD,Levin:2011C,2014PhRvD..90h4016H,habhu:2019,2013PhRvD..87l7501H,Yunes:2009,Habib:2020dba,MC:2015PhRvD,Moore:2016,Tai:2014,Will:2012,Tanay:2016,sam:2017ApJ,Gayathri:2020coq,hoang:2017APJ,gondkoc:2018G,gon:2017,2021PhRvD.103f4022I,2021PhRvD.103h4018C,Osburn:2016}. 
Strides in the modeling and understanding of 
eccentric compact binary mergers has been accompanied 
by population synthesis models~\cite{rocarl:2018PhRvDR,samjoh:2018Z,sam:2018PhRvD3014S,PhysRevD.97.103014} that have been 
significantly improved to be compatible with 
the observation of 
stellar mass black holes in dense 
stellar environments, such as globular clusters 
in our galaxy~\cite{cho:2013ApJ,2012Natur49071S,2021NatAs...5..957G}, 
and galactic nuclei~\cite{Kocsis:2012,Leary:2009,Leigh:2018MNRAS}.

Impelled by these theoretical and observational 
advances, researchers have developed the required tools to 
search for this astrophysical population in gravitational 
wave data~\cite{Tiwari:2016,Romero-Shaw:2019itr,Nitz:2019spj,Adam:2018prd,Salemi:2019owp,2020arXiv201203963W}. Some recent 
studies have attempted to constrain the eccentricity 
of actual gravitational wave sources~\cite{2022arXiv220614695R}. 
A plethora of studies for the massive stellar black hole merger 
named GW190521~\cite{Abbott:2020tfl} provide persuasive 
evidence for the 
existence of eccentric compact binary mergers in dense 
stellar environments~\cite{2020ApJ...903L...5R,Samsing2022AGNAP,Gayathri:2020coq}. It is expected that several tens of eccentric 
compact binary mergers observed by advanced ground-based gravitational wave detectors will suffice to 
understand what formation channels contribute or dominate the eccentric merger rate~\cite{2022arXiv220614695R}.

In view of these developments, 
and the upcoming deluge of gravitational wave observations to 
be enabled by advanced LIGO~\cite{LSC:2015,LIGOScientific:2016dsl} and its international counterparts VIRGO and KAGRA~\cite{Virgo:2015,Salemi:2019owp,kagra_2021},
it is timely and relevant to continue developing adequate 
tools for the identification of gravitational wave signals 
that may be produced by eccentric compact binary mergers.

The best tool at hand to gain insights about the physics of 
eccentric binary black hole mergers is numerical relativity, 
and thus we use the open source, community-driven, numerical 
relativity software, the Einstein Toolkit~\cite{ETL:2012CQGra} 
to produce a suite 
of numerical relativity waveforms that describe eccentric, 
spinning, nonprecessing binary black hole mergers. Non-spinning, eccentric simulations were investigated in previous works in~\cite{Adam:2018prd,huerta_nr_catalog}. These 
waveforms include higher order modes up to \(\ell\leq4\), except 
for memory modes. We use these numerical relativity waveforms 
to carry out the following studies:

\begin{itemize}[nosep]
    \item \textbf{Gravitational wave detection} We 
    construct two types of waveforms that include either
    quadrupole modes, \(\ell=|m|=2\), or modes up to 
    \(\ell\leq4\). We assume stellar mass binary black holes 
    that may be observed by advanced LIGO-type detectors and 
    compute signal-to-noise ratio (SNR) calculations for a 
    variety of astrophysical scenarios, and explore 
    whether the inclusion of higher order wave modes leads to 
    measurable SNR increases. 
    \item \textbf{Gravitational wave modeling} We use two effective-one-body (EOB) eccentric waveform models: \texttt{TEOBResumS}~\cite{damour_2014,nagar_2016,nagar_2018,nagar_2020a,nagar_2020b,riemenschneider_2021,chiaramello_nagar_2020,nagar_2021} and \texttt{SEOBNRE}~\cite{cao:2017,2020PhRvD.101d4049L,liu_higher-multipole_2022} to estimate the eccentricities of our 
    numerical relativity waveforms.
    This exercise was 
    useful to identify areas of improvement for next generation 
    waveform models, and to get a better understanding of signals 
    that may be discovered in upcoming gravitational wave searches. Note that due to conventions and different definitions of eccentricity, the inferred eccentricities cannot be directly compared with each other. A detailed comparison between the two waveform models is given in \citet{knee_rosetta_2022}.
    \item \textbf{Parameter space degeneracy} We quantified 
    the impact of including higher order modes in terms 
    of fitting factor calculations that aim to pinpoint 
    an optimal quasicircular NRHybSur3dq8 waveform 
    signal~\cite{nrhybsur3dq8_2019} whose astrophysical 
    parameters best reproduce the complex morphology of moderately 
    or highly eccentric numerical relativity waveforms. 
\end{itemize}

\noindent These three complementary studies underscore 
the importance of improving our understanding of 
compact binary mergers in dense stellar environments. 
It is not enough to hope for the best and expect that burst 
or machine learning searches identify complex signals 
in gravitational wave data~\cite{Adam:2018prd,2020arXiv201203963W}. 
It is also necessary to develop a comprehensive toolkit 
that encompasses numerical relativity waveforms, semi-analytical or 
machine learning based models, 
and signal processing tools to detect and then infer the 
astrophysical properties of eccentric compact binary mergers. 
Not doing so would be a disservice to 
the proven detection capabilities of advanced gravitational 
wave detectors, and would limit the science reach of 
gravitational wave astrophysics. To contribute to this 
important endeavor, we release our catalog of 
numerical relativity waveforms along with this article.

This article is organized as follows. We describe our 
approach to create a catalog of eccentric numerical 
relativity waveforms in Sec.~\ref{sec:set_up}. Sec.~\ref{sec:snr_calcs} presents our waveform 
catalog, and a systematic study on the importance of including 
higher order wave modes in terms of SNR calculations. In  Sec.~\ref{sec:par_deg} we study whether surrogate models 
based on quasicircular, spinning, nonprecessing binary 
black hole numerical relativity waveforms can capture the 
physics of spinning, nonprecessing eccentric mergers.
We summarize our findings and future directions of work in 
Sec.~\ref{sec:end}.

\section{Numerical Setup and Simulation Details}
\label{sec:set_up}

\begin{figure*}
    \centering
    \includegraphics[height=\textheight]{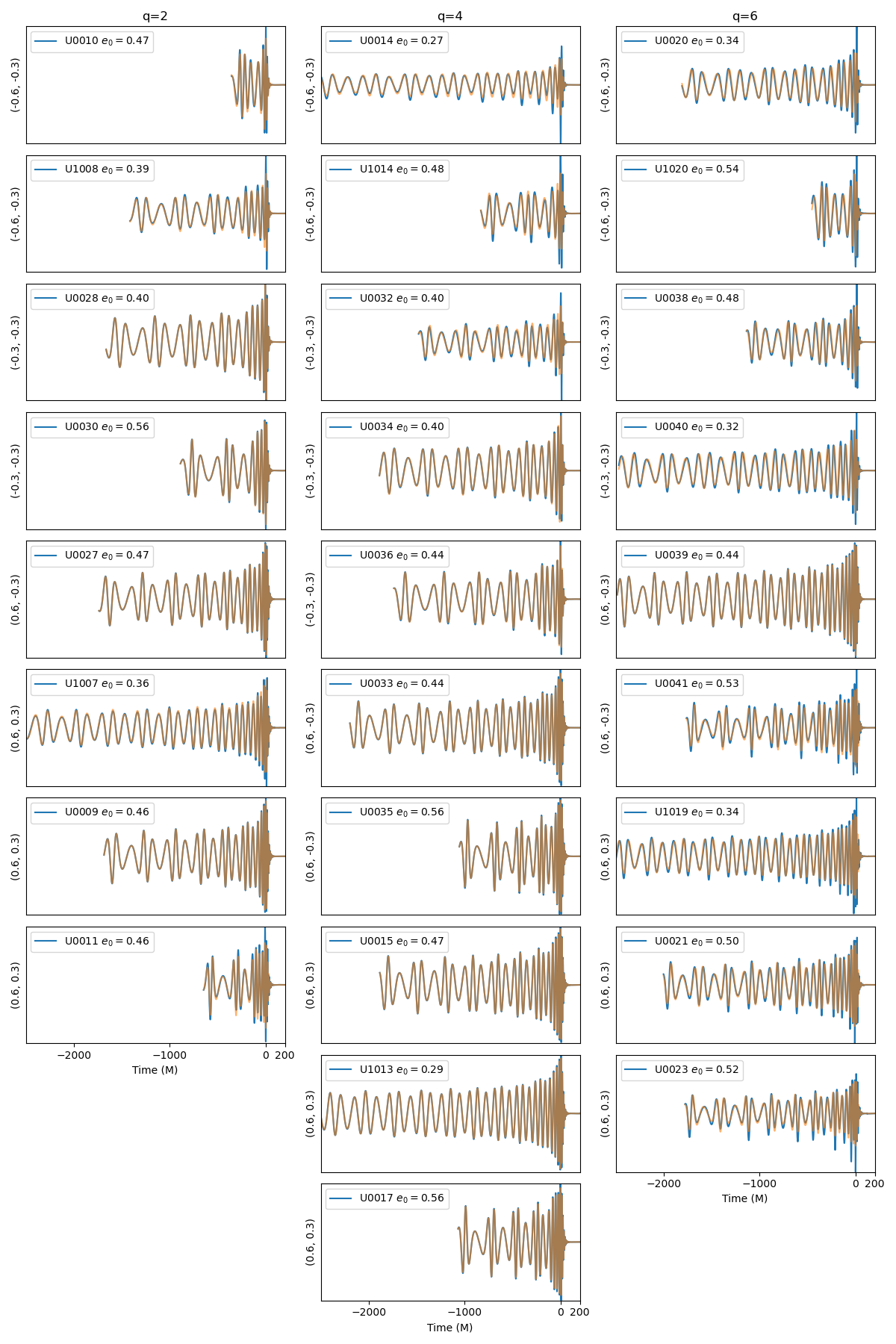}
    \caption{\textbf{Numerical relativity waveform catalog} Each column is associated with a 
    given mass-ratio \(q=\{2, 4, 6\}\). 
    From top to bottom, simulations are ordered in $(\chi_{1z},\chi_{2z})$. The eccentricity $e_0$ inferred from \texttt{TEOBResumS} is given in the label. Each panel presents two types of waveforms: a \(\ell=|m|=2\) signal (orange), and 
    one that includes higher 
    order modes (blue). We have selected the inclination 
    of the binary that maximizes the contribution higher order modes.}
    \label{fig:gallery}
\end{figure*}

 We used the \texttt{Einstein Toolkit} to generate a 
 catalog of numerical relativity waveforms. Initial data for the binaries was computed using the \texttt{TwoPunctures} code. The evolution was done with the CTGamma code implementing the 3+1 BSSN formulation. The outer boundary of the simulation domain was placed far enough (2500M) to avoid any contamination of the signal until 200M after the merger. Each simulation was run at three resolutions to check for convergence (see appendix \ref{app:convergence}): $N=36,40,44$ where $N$ is the resolution across the finest grid radius. The highest resolution simulations were used for all analyses. Further details of the simulation setup are given in \cite{huerta_nr_catalog}. Waveforms extracted at future null infinity were computed for $1<l\leq 4$ and $1 \leq |m|\leq l$ modes using the POWER code \cite{johnson_power} by extrapolating the observed signals from 7 detectors located 100\textendash 700M. $m=0$ modes were not used since these modes (so-called memory modes) are many orders of magnitude smaller than the dominant modes of the waveform making a reliable estimation difficult due to numerical resolution (for more details see Sec. 6.2 in \cite{favata_gravitationalwave_2010}). A plot of all the $h_+$ simulation waveforms is shown in Fig.~\ref{fig:gallery}. Note that the simulations are also dimensionalized in units of M.
 
Table~\ref{table:simulations} describes the properties of 
our waveform catalog, including the mass-ratio, 
individual spins and orbital eccentricity of each binary (measured from both waveform templates). 
The library consists of 27 simulations across 3 mass ratios,
\(q = \{2,4,6\}\), and a combination of nonprecessing 
individual spins, namely $\pm 0.6$ and $\pm 0.3$, for 
the primary (heavier) and secondary (lighter) 
binary components, respectively.

\section{Eccentricity Measurements}\label{sec:ecc_measurement}
Orbital eccentricity in a Keplerian interpretation can only be defined for a BBH system during the early inspiral, where the orbits of the binaries are nearly closed (the adiabatic approximation). This definition breaks down close to the merger, which is when our simulations begin. Thus, the definitions of eccentricity used to generate the initial conditions are ill-defined, even though they produce eccentric simulations.

Using evolution information of the binary, such as the separation between the components, throughout the simulation to obtain a measure of orbital eccentricity is not useful, as such a concept is gauge-dependent by assuming a coordinate system. To obtain a useful measure of eccentricity, we calibrate our numerical simulations to the spin-aligned eccentric EOB models \texttt{TEOBResumS} and \texttt{SEOBNRE}. For both of these models, a reference eccentricity $e_0$ and reference GW frequency $f_\mathrm{ref}$ are used as inputs to generate adiabatic initial conditions of the binary from which the waveform is computed. As investigated in~\citet{knee_rosetta_2022}, each waveform model's definition of $e_0$ may vary, due to different conventions of $f_\mathrm{ref}$ and initial condition constructions.


The method is similar to that used in~\cite{2020PhRvD.101d4049L,habhu:2019}. The key idea consists of using 
\(\ell=|m|=2\) waveforms to compute 
the fitting factor between a given numerical 
relativity waveform, and an array of templates.
In this work, we have assigned $f_\mathrm{ref}=10\mathrm{Hz}$, which is at the lower end of the detectability range for LIGO.
To estimate the eccentricity of our 
numerical relativity waveforms, we need to 
compute a few objects. The first of them 
is the inner product between 
one of our numerical relativity waveforms, \(h_{22}^{\text{NR}}\), 
and a waveform template, 
\(h_{22}^{\text{template}}\), given by:

\begin{equation}
    \langle h_{22}^{\text{NR}} | h_{22}^{\text{template}} \rangle = \mathcal{R} \left[ \int_{t_1}^{t_2} h_{22}^{\text{NR}} h_{22}^{\text{* template}}  \right]\,.
\end{equation}

\noindent Where $\mathcal{R}$ represents the real component. Note that the inner product is calculated by maximizing over both the time and phase of the two waveforms. $t_1$  represents an initial time at a 
point free from initial junk radiation, 
and $t_2$ marks the end of 
the numerical relativity simulation. $t_2$ in general is 50\textendash 100M after merger for the signal to reach the outermost detectors in the simulation, but not long enough so that the initial junk radiation gets reflected back to the detectors due to the outer Dirichlet boundary conditions.
The norm of a waveform is given by:

\begin{equation}
    ||h|| \equiv \sqrt{\langle h | h \rangle}\,.
\end{equation}

\noindent With these two quantities, we can 
compute the fitting factor between one of our 
numerical relativity waveforms and a bank of 
waveform templates, and thus measure the eccentricity $e_0$ as:

\begin{equation}\label{eq:fittingfactor}
    \mathrm{FF} \equiv \max_{t_0,\phi_0} \frac{\langle h_{22}^{\text{NR}} | h_{22}^{\text{template}} \rangle}{||h_{22}^{\text{NR}}|| \cdot ||h_{22}^{\text{template}}||}\,,
\end{equation}

\begin{equation}
    e_0 = \argmax_{e_0} \left( \mathrm{FF} \right)\,,
\end{equation}

\noindent where the eccentricity $e_0$ is defined at 
the lower frequency bound $f_{\text{ref}}$ 
which determines the length of the simulation 
prior to merger for the template. This calculation 
essentially corresponds to the inner product of 
a numerical relativity waveform maximized over a 
bank of \text{SEOBNRE} and 
\text{EOBResumS} templates. Note that to dimensionalize $f_\mathrm{ref}$, the total mass of the binary ($M$) needs to be provided. Thus, all inferences of eccentricity are dependent on the choice of $M$, which differed based on the waveform template code for stability purposes.

All inferred eccentricities for both \texttt{TEOBResumS} and \texttt{SEOBNRE} are given in Table \ref{table:simulations}.

\subsection{\texttt{TEOBResumS} inferences}
For \texttt{TEOBResumS} waveforms (produced 
with the \texttt{TEOBResumS-DALI} branch), we set the total mass of the binary system $M=30M_\odot$ and $f_\mathrm{ref}=10\mathrm{Hz}$. Scans were made up to $e_0=0.8$, with a resolution of 0.001. To note, for \texttt{TEOBResumS}, the waveform begins from apastron and while we maximize the fitting factor by changing the initial phase $\phi_0$, this resulting definition of $f_\mathrm{ref}$ is different from that used in \texttt{SEOBNRE}.

\begin{figure*}
    \centering
    \includegraphics[width=0.95\linewidth]{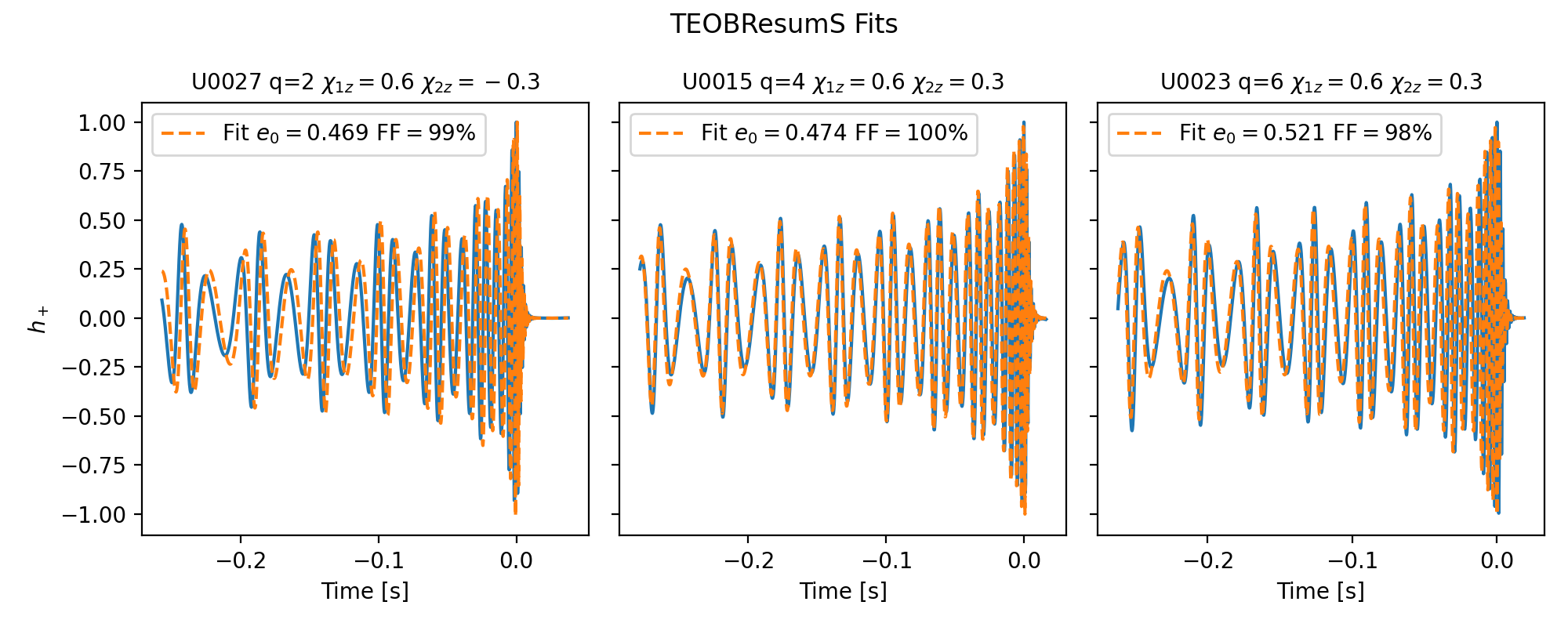}
    \caption{\textbf{Comparison between numerical relativity waveforms and \texttt{TEOBResumS}.} Comparison of three waveforms overlaid with the best matching \texttt{TEOBResumS} waveform, effectively calibrating the eccentricity $e_0$ of the waveform. The total mass of the binary is $M=30M_{\odot}$ and the reference frequency is $f_\mathrm{ref}=10\mathrm{Hz}$. Solid lines represent numerical relativity 
    waveforms, while dotted lines represent optimal \texttt{TEOBResumS} templates.}
    \label{fig:teob_comp}
\end{figure*}

From Table \ref{table:simulations}, we see that good matches are obtained for nearly all the simulations\textemdash 23 out of 27 simulations have $\mathrm{FF}>90\%$. The remaining simulations that do not match well visually appear to be of high eccentricity (possibly $e_0>0.8$) which would be beyond the explored parameter space. Fig. \ref{fig:teob_comp} shows a comparison between the simulations and the best fitting \texttt{TEOBResumS} for 3 simulations.

\subsection{\texttt{SEOBNRE} inferences}

To produce this bank of \texttt{SEOBNRE} templates, 
we set $f_{\text{ref}}=10\mathrm{Hz}$. 
To obtain stable \texttt{SEOBNRE} waveforms, we set the 
total mass of the binary system $M=60M_\odot$ 
for $e\leq0.5$ and $M=30M_\odot$ for $e>0.5$. 
Higher mass binaries spend less cycles in the 
detectable frequency band, and so for highly 
eccentric simulations, the code does not have 
enough inspiral points to produce an accurate 
waveform, requiring a smaller mass for stability. Lower mass binaries at low eccentricities produced waveforms that were too large, and thus $M=60M_\cdot$ was chosen for efficiency.

As seen in Table~\ref{table:simulations}, 
we find good fitting factors for roughly half of the simulations. For some highly eccentric simulations, a suitable match was not found. This is because some numerical 
relativity waveforms contain moderately spinning binaries with highly eccentric orbits that are beyond the realm of 
applicability of the SEOBNRE model. It is 
possible to quantify the reliability of SEOBNRE 
signals with the 
``spin hang-up parameter'', $\chi_\mathrm{up}$~\cite{2020PhRvD.101d4049L}

\begin{equation}
    \chi_\mathrm{up} = \frac{8 \chi_\mathrm{eff} + 3 \sqrt{1 - 4 \eta}\chi_A}{11}\,,
\end{equation}

\noindent where $\chi_\mathrm{eff} = (q\chi_{1z} + \chi_{2z})/(1+q)$, $\chi_A = (q\chi_{1z} - \chi_{2z})/(1+q)$ and $\eta = m_1 m_2 / M^2$ for a binary of masses $(m_1,m_2)$ and (orbit aligned) dimensionless spins $(\chi_{1z}, \chi_{2z})$ respectively. 
Furthermore, $M=m_1 + m_2$ and $q = m_1/m_2 \geq 1$.

For simulations with poor matches, we find that 
$\chi_\mathrm{up} > 0.35$, and visual inspection of 
these waveforms suggest high eccentricity, $e_0>0.6$. The SEOBNRE template waveform is inaccurate in producing reliable waveforms in that region of parameter space~\cite{2020PhRvD.101d4049L}. 
Indeed, for the two simulations with $\chi_{up}=-0.5$, we were unable to obtain a suitable waveform. Nevertheless, in the valid regions, eccentricities are found to good accuracy. For simulations with $\text{FF}<75\%$ the eccentricity is considered unconstrained, and we simply report the best match for completeness.

\begin{table}[!ht]
    \centering
    \begin{tabular}{l|l|l|l|l|l|l}
        \thead{Simulation} & \thead{q} & \thead{$\chi_{1z}$} & \thead{$\chi_{2z}$} & \thead{\texttt{TEOBResumS} \\ $e_0$(FF)} & \thead{\texttt{SEOBNRE} \\ $e_0$(FF)} & \thead{$\chi_{\mathrm{up}}$}\\ \hline \hline
        U1007 & 2 & 0.6 & 0.3 & 0.36 (97.2\%) & 0.39 (94.8\%) & 0.39 \\ \hline
        U1008 & 2 & -0.6 & -0.3 & 0.39 (98.3\%) & 0.67 (94.0\%) & -0.39 \\ \hline
        U0009 & 2 & 0.6 & 0.3 & 0.46 (97.2\%) & 0.70 (29.3\%) & 0.39 \\ \hline
        U0010 & 2 & -0.6 & -0.3 & 0.47 (88.5\%) & 0.79 (99.3\%) & -0.39 \\ \hline
        U0011 & 2 & 0.6 & 0.3 & 0.46 (55.6\%) & 0.08 (45.7\%) & 0.39 \\ \hline
        U0027 & 2 & 0.6 & -0.3 & 0.47 (99.0\%) & 0.70 (90.3\%) & 0.26 \\ \hline
        U0028 & 2 & -0.3 & -0.3 & 0.40 (98.7\%) & 0.66 (96.7\%) & -0.23  \\ \hline
        U0030 & 2 & -0.3 & -0.3 & 0.56 (76.7\%) & 0.77 (84.0\%) & -0.23  \\ \hline
        & & & & \\ \hline
        U0014 & 4 & -0.6 & -0.3 & 0.27 (99.4\%) & 0.26 (99.8\%) & -0.46 \\ \hline
        U1013 & 4 & 0.6 & 0.3 & 0.29 (97.5\%) & 0.33 (61.2\%) & 0.46 \\ \hline
        U1014 & 4 & -0.6 & -0.3 & 0.48 (98.3\%) & 0.68 (95.0\%) & -0.46 \\ \hline
        U0015 & 4 & 0.6 & 0.3 & 0.47 (99.5\%) & 0.05 (28.7\%) & 0.46 \\ \hline
        U0017 & 4 & 0.6 & 0.3 & 0.56 (93.4\%) & 0.45 (30.2\%) & 0.46 \\ \hline
        U0032 & 4 & -0.3 & -0.3 & 0.40 (99.2\%) & 0.40 (87.2\%) & -0.25 \\ \hline
        U0033 & 4 & 0.6 & -0.3 & 0.44 (97.6\%) & 0.69 (18.6\%) & 0.39 \\ \hline
        U0034 & 4 & -0.3 & -0.3 & 0.40 (98.7\%) & 0.41 (93.1\%) & -0.25 \\ \hline
        U0035 & 4 & 0.6 & -0.3 & 0.56 (79.2\%) & 0.49 (25.7\%) & 0.39   \\ \hline
        U0036 & 4 & -0.3 & -0.3 & 0.44 (91.7\%) & 0.70 (79.7\%) & -0.25 \\ \hline
        & & & & \\ \hline
        U0020 & 6 & -0.6 & -0.3 & 0.34 (99.7\%) & N/A & -0.50 \\ \hline
        U1019 & 6 & 0.6 & 0.3 & 0.34 (94.9\%) & 0.58 (12.7\%) & 0.50 \\ \hline
        U1020 & 6 & -0.6 & -0.3 & 0.54 (93.8\%) & N/A & -0.5 \\ \hline
        U0021 & 6 & 0.6 & 0.3 & 0.50 (93.7\%) & 0.20 (15.6\%) & 0.50 \\ \hline
        U0023 & 6 & 0.6 & 0.3 & 0.52 (98.4\%) & 0.60 (13.1\%) & 0.50 \\ \hline
        U0038 & 6 & -0.3 & -0.3 & 0.48 (97.3\%) & 0.70 (94.1\%) & -0.26 \\ \hline
        U0039 & 6 & 0.6 & -0.3 & 0.44 (96.2\%) & 0.26 (29.8\%) & 0.45 \\ \hline
        U0040 & 6 & -0.3 & -0.3 & 0.32 (98.3\%) & 0.41 (96.2\%) & -0.26 \\ \hline
        U0041 & 6 & 0.6 & -0.3 & 0.53 (96.8\%) & 0.47 (13.6\%) & 0.45 \\
    \end{tabular}
    \caption{\textbf{Physical parameters of numerical relativity waveform catalog} Mass-ratio, \(q\), 
    individual spins, $(\chi_{1z}, \chi_{2z})$, and 
    estimated orbital eccentricity, $e_0$, of 
    our numerical relativity waveforms.}
    \label{table:simulations}
\end{table}

\subsection{Comparison of the two waveform models}
For a detailed comparison between the two waveform models, we refer to \citet{knee_rosetta_2022} which goes into detail about the systematic differences. Two results that can be corroborated is the fact that the \texttt{TEOBResumS} calibrated $e_0$ is uniformly less than that of \texttt{SEOBNRE} ($e_0^\texttt{TEOB} < e_0^\texttt{SEOBNRE}$). Moreover, the disparity is low at $e_0^\texttt{TEOB}\approx0.2$ and increases up to $50\%$ for higher eccentricities.

\section{Importance of higher order harmonics}
\label{sec:snr_calcs}

Having computed higher order wave modes, $h^{lm}(t)$, 
we can construct the full waveform 

\begin{equation}\label{eq:waveform_decomp}
    h(t,\theta,\phi) = h_+ + ih_x = \sum_{l \geq 2} \sum_{m\geq -l}^{m\leq l} h^{lm} {}_{-2}Y_{lm}(\theta,\phi)\,,
\end{equation}

\noindent where ${}_{-2}Y_{lm}(\theta,\phi)$ are the spin-weight\textendash 2 spherical harmonics computed at a particular inclination ($\theta$) and azimuth ($\phi$). $\theta=0$ corresponds to observing the binary face-on i.e., with the orbital angular momentum vector pointed toward the observer. 

Since nearly eccentric waveforms resemble 
quasicircular signals near merger due to 
circularization, 
we compute the importance of including 
higher order harmonics on 
the signal across the entire waveform evolution to better quantify the effect of eccentricity. From the results of \cite{Adam:2018prd} Sec. III, the $\Delta \mathcal{B}$ metric is used. It involves integrating over the entire numerical relativity waveform (after removing junk radiation)

\begin{equation}
    \mathcal{B}^{(l,|m|)}(\theta,\phi) = \int_{t=t_0}^T \sqrt{h(t,\theta,\phi)\Tilde{h}(t,\theta,\phi)}\mathrm{d}t\,,
\end{equation}

\begin{equation}\label{eqn:delta_b}
    \Delta \mathcal{B}(\theta,\phi) = \frac{\mathcal{B}^{(\ell,|m|)}(\theta,\phi) - \mathcal{B}^{(\ell=|m|=2)}(\theta,\phi)}{\mathcal{B}^{(\ell=|m|=2)}(\hat{\theta},\hat{\phi})}\,,
\end{equation}

\noindent where $(\hat{\theta},\hat{\phi})$ represent the orientation that maximizes the $(\ell=|m|=2)$ mode of $\mathcal{B}$. To find the $(\theta,\phi)$ 
combination that 
maximizes the contribution of higher order 
modes in terms of SNR calculations, we scan across $(\theta,\phi)$ space at a resolution of $0.01$ radians and select the orientation ($\theta^*,\phi^*$) that maximizes $\Delta \mathcal{B}$ in Eq.~\eqref{eqn:delta_b}. The resultant optimal orientation is usually within three categories: one with the inclination close to the pole, one with inclination close to the equator and one slightly apart from both these angles.

To quantify the impact higher order modes would have on ground based detectors, we focus on the optimal SNR response of a waveform [$\mathbf{h}$] as \cite{Adam:2018prd}
\begin{equation}
    \mathrm{SNR}[\mathbf{h}]^2 = 4 \mathcal{R}\int_0^\infty \frac{\Tilde{h}(f)\Tilde{h}^*(f)}{S_n(f)}\mathrm{d}f\,,
\end{equation}

\noindent where $S_n(f)$ is the one-sided power spectral density (PSD) for LIGO's Zero Detuned High Power configuration (ZDHP) \cite{ZDHP:2018}. We thus compute  SNRs for both $(\ell,|m|)$ and $(\ell=|m|=2)$ modes across all sky locations $(\alpha,\beta)$ with the optimized orientation $(\theta^*,\phi^*)$. For the following 
results we set the polarization angle 
to $\psi=\pi/4$, and compute the effect of the higher order modes as

\begin{equation}
    \Delta \mathrm{SNR} = \frac{\mathrm{SNR}^{(\ell,|m|)} - \mathrm{SNR}^{(\ell=|m|=2)}}{\mathrm{\hat{SNR}}^{(\ell=|m|=2)}}\,,
    \label{eq:delta_snr}
\end{equation}

\noindent where $\mathrm{\hat{SNR}}^{(\ell=|m|=2)}$ is the 
maximum value of the $\ell=|m|=2$ mode across the 
sky $(\hat{\alpha},\hat{\beta})$. The total mass 
of the binary is set to \(M=60M_{\odot}\).

\begin{figure}[h]
    \centering
    \includegraphics[width=0.9\linewidth]{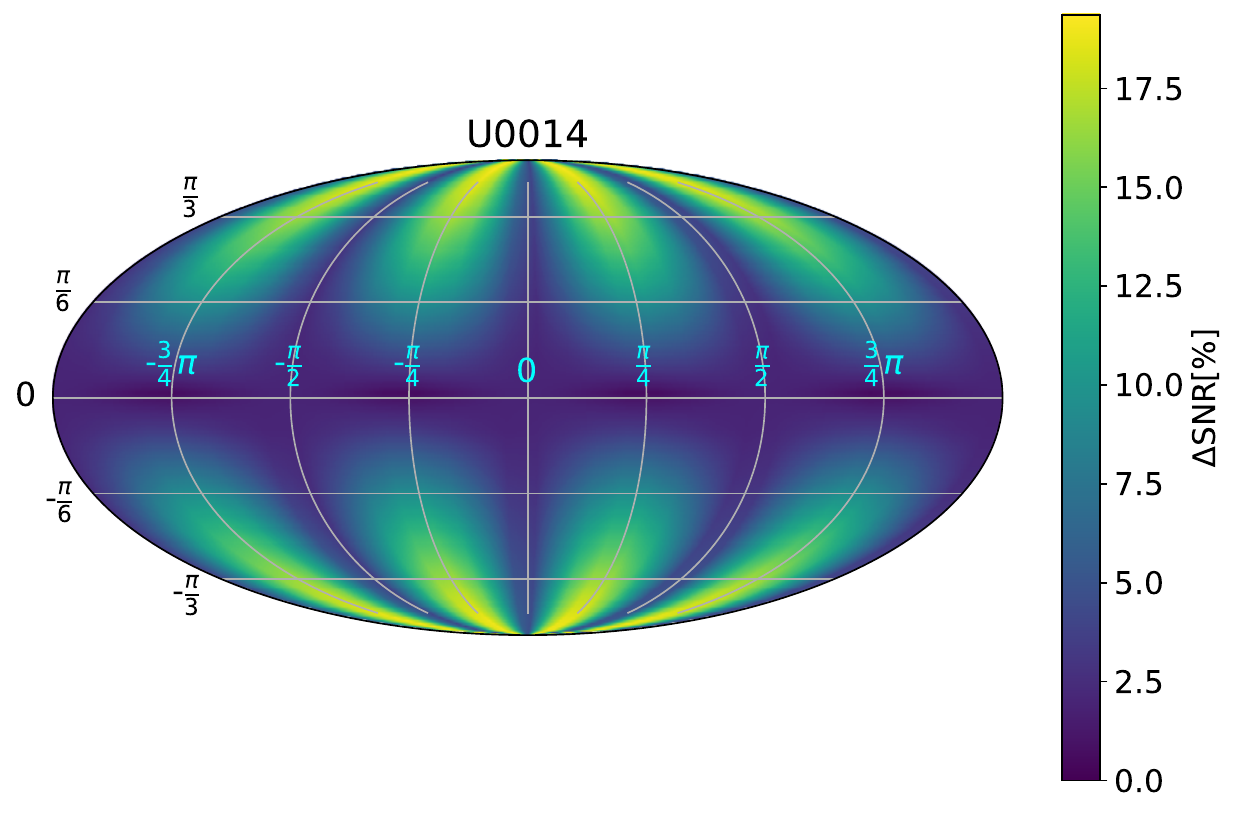}
    \includegraphics[width=0.9\linewidth]{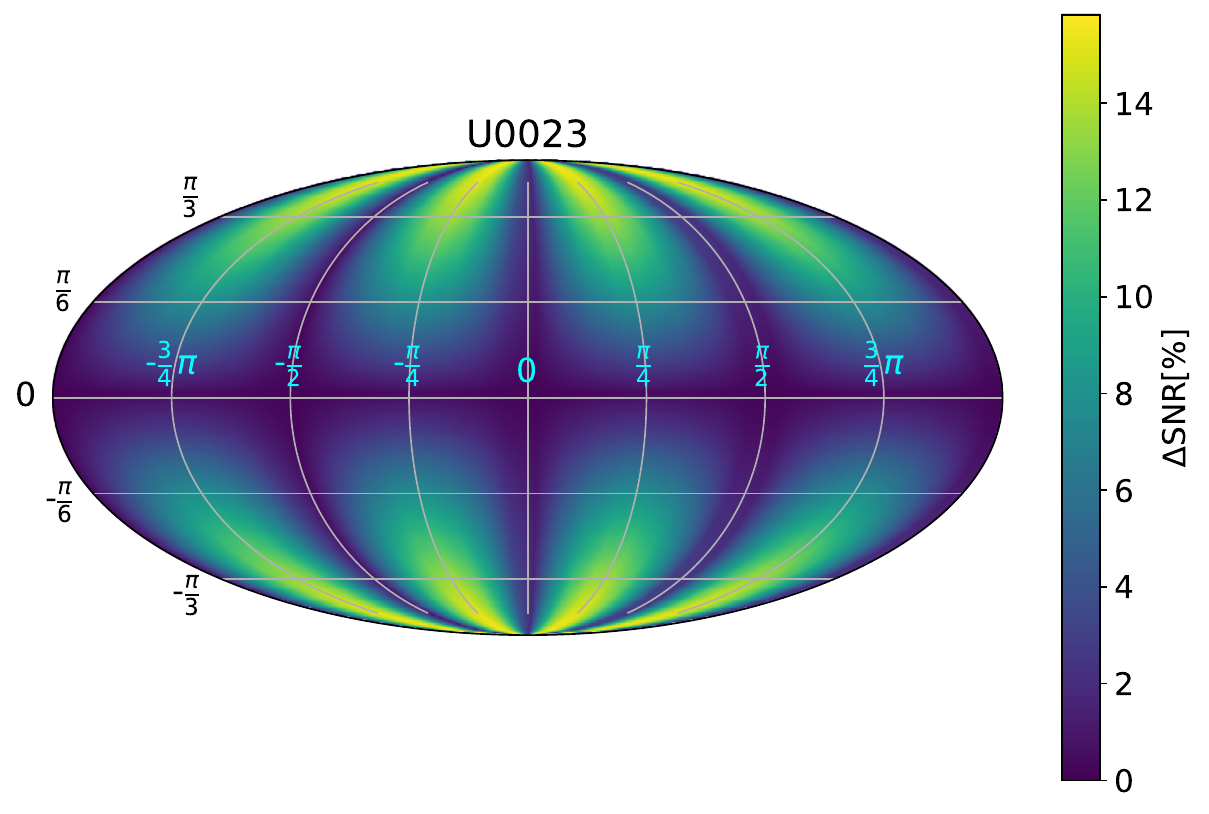}
    \caption{\textbf{Importance of higher order modes for SNR calculations} The panels show the high increase in SNR, \(\Delta \textrm{SNR}\) in Eq.~\eqref{eq:delta_snr}, as a result of including higher order modes in the modeling of eccentric, spinning, binary black hole mergers. We assume an advanced LIGO-type detector, and binaries with total mass \(M=60M_{\odot}\) for numerical relativity waveform 
    U0014 (top panel) and U0023 (bottom panel).}
    \label{fig:snr_comp}
\end{figure}

The results can be categorized into three different 
categories depending on what the optimal orientation of the binary is  $(\theta^*,\phi^*)$. The first category is that in which 
$\theta^*\rightarrow 0$, and the inclusion of higher order 
modes has a marginal impact on the SNR of the signal, 
typically no more than 4\%. The second category is for 
$70^\circ < \theta^* < 110^\circ$, in which case the 
contribution of 
higher order modes to the SNR of the signal is 
significant, with \(\Delta \textrm{SNR} \sim 25\%\). 
The final category 
are the in-between values of $\theta^*$ for which 
$\Delta \mathrm{SNR}$ will be intermediate to that 
of the first two categories. Fig. \ref{fig:snr_comp} shows the high effect of higher order modes on the skymap for two simulations. Increasing the mass of the binary to $M=80M_\odot$ yields an increase of SNR to nearly 25\% for some of the simulations.

These studies underscore the importance 
of including higher order modes in the modeling 
and detection of eccentric compact binary mergers, 
since SNR increases of order 
\(\Delta \textrm{SNR}\sim 20\%\) mean that marginally detectable signals~\cite{Brown_2013} 
may then become easier to detect, or observable to 
larger distances.

\section{Comparisons with quasicircular waveforms}
\label{sec:par_deg}

Studies in the literature have shown that the morphology 
of non-spinning, mildly eccentric binary black hole mergers 
may be captured by quasicircular, spinning, nonprecessing 
binary black hole mergers~\cite{ENIGMA_Huerta}. 
Here we quantify whether this parameter space degeneracy 
between orbital eccentricity and spin corrections still 
remain when we directly compare our new set of eccentric, 
spinning, nonprecessing numerical 
relativity waveforms with the \texttt{NRHybSur2dq8} 
surrogate model~\cite{PhysRevD.99.064045} 
that describes quasicircular, 
spinning, nonprecessing mergers.

\begin{figure*}
    \centering
    \includegraphics[width=.95\textwidth]{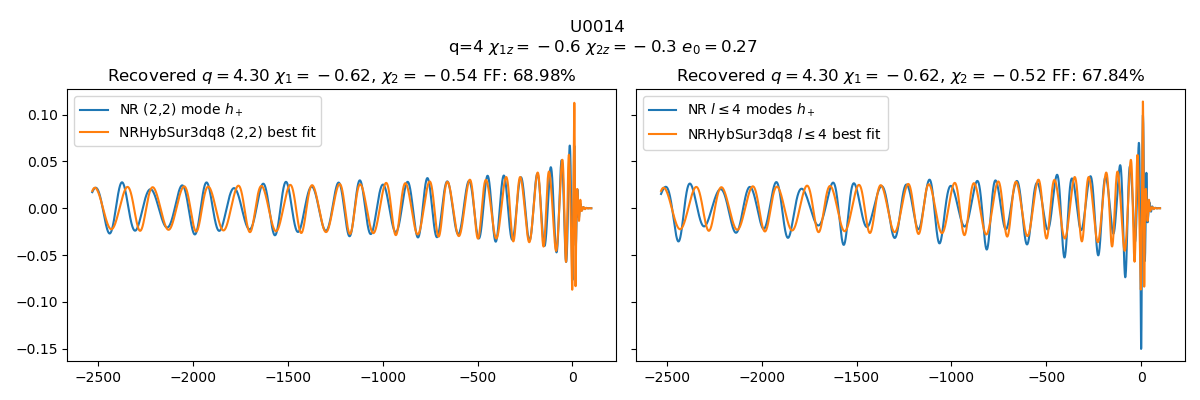}
    \includegraphics[width=.95\textwidth]{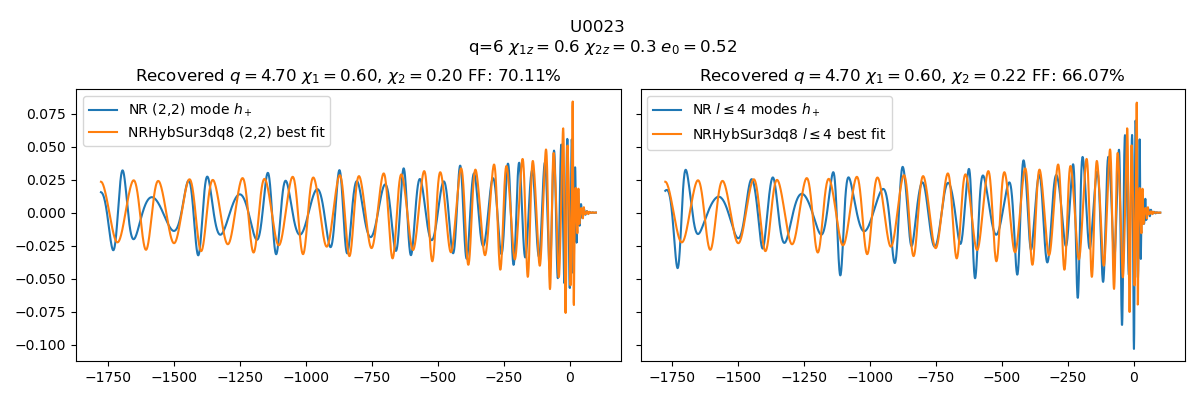}
    \caption{\textbf{Non parameter space degeneracy between spin and eccentricity corrections} Fitting factor (FF) calculations 
    between numerical relativity waveform U0014 (top panels) and 
    U0023 (bottom panels) and NRHybSur2dq8 waveform templates. 
    In both cases, we show results for signals that include only \(\ell=|m|=2\) modes (left panels) and higher order modes (right panels). We notice significant discrepancies between ground-truth and recovered values for the mass-ratio and 
    individual spins of the binary components through FF calculations.}
    \label{fig:degeneracy_U0015}
\end{figure*}

We carry out this study by computing fitting factor 
calculations, see Eq.~\eqref{eq:fittingfactor}, 
between a given waveform in our numerical relativity 
catalog and an array of \texttt{NRHybSur2dq8} waveforms 
that scan the \((q, \chi_{1z}, \chi_{2z})\) 
parameter space using a simple grid search. We use 
an interval of size \(\Delta q = 2\) centered around the 
truth mass-ratio.
So for numerical relativity waveforms of 
mass-ratio \(q=4\), we scan an interval that covers 
the range \(2\leq q \leq 6\) (note for \(q=2\) the interval is
\(1\leq q \leq 4\)). For individual spins, we 
consider the range $-0.7\leq \chi_{\{1z,2z\}} \leq  0.7$. 
The resolution of the search is \(\delta q = 0.1\), 
and \(\delta \chi = 0.02\) for both spins. 
Following these conventions, we consider two cases. 
In the first both numerical 
relativity waveforms and \texttt{NRHybSur2dq8} waveforms 
include only the \(\ell=|m|=2\) mode, whereas in the second 
case both types of waveforms include higher order modes. 
Results of this analysis for simulations U0014 and U0023 
are presented in Fig.~\ref{fig:degeneracy_U0015}.

Additional results for other numerical simulations in our 
waveform catalog may be found in 
Table~\ref{tab:degeneracy_params}. These findings, 
along with the results we presented in 
Table~\ref{table:simulations} using the SEOBNRE 
waveform family, exhibit the importance of 
developing waveform models that are informed by 
numerical relativity simulations to accurately capture 
orbital eccentricity and spin corrections. At this time, 
these results show that moderately or highly eccentric and 
spinning signals may not be captured by template matching 
algorithms, unless the signal is loud enough to be 
captured by unmodeled (burst) searches.

In summary, this study shows that it is not possible 
for quasicircular, 
spinning, nonprecessing signals to capture the dynamics of 
moderately and highly eccentric, spinning, nonprecessing 
signals. We either develop the required methods (waveforms \& 
signal processing tools) to search for and find these signals 
or we may miss an interesting population of compact 
binary sources. 

\section{Conclusions}
\label{sec:end}

We have presented a set of 27 eccentric, spin-aligned 
binary black hole simulations that describe three 
different mass-ratios \(q=\{2, 4, 6\}\). To measure 
the eccentricity of the simulations, we computed 
fitting factors against two spin-aligned eccentric effective-one-body models with eccentricity\textemdash \texttt{TEOBResumS} and \texttt{SEOBNRE}.
We were able to estimate eccentricities for nearly all of the simulations with \texttt{TEOBResumS}, with eccentricity ranges of $0.27\leq e_0^\texttt{TEOB}< 0.58$ and roughly half of 
the simulations with \texttt{SEOBNRE} with eccentricity ranges $0.26\leq e_0^\texttt{SEOBNRE}<0.8$. The remaining simulations appear to be of even 
higher eccentricity, though producing such waveforms 
from templates proved to be difficult for the values of 
spins and orbital eccentricities used in our simulations. Current limitations to the 
existing SEOBNRE library will be alleviated by including 
higher order eccentricity terms, which become 
increasingly important at 
higher mass ratios as indicated by our findings and 
those reported in~\cite{2020PhRvD.101d4049L}. Indeed in~\citet{liu_higher-multipole_2022}, the authors introduce a new model \texttt{SEOBNREHM} that utilizes these higher order terms greatly that improves fitting factors and produces accurate waveforms for maximally spinning, highly eccentric simulations. Comparing our simulations with this model is a future project that may yield new results.

For these simulations, we 
performed the following analyses:

\begin{enumerate}
    \item Selecting the orientation of the binary that maximizes the contribution of higher order modes, we computed the SNR observed for ground-based LIGO-type detectors across the sky. In doing so, we observed that for simulations, the inclusion 
    of high order modes in the waveform increases the SNR between 5\textendash 35\%.
    \item We do not find significant parameter space 
    degeneracies between spinning, eccentric waveforms and 
    quasicircular, spinning waveforms upon 
    computing fitting factor calculations assuming a coarse grid search across mass ratio, and spins. In general the fitting factors are worse when comparing higher order modes.
\end{enumerate}

These analyses underscore the importance of 
using numerical relativity to understand the physics of 
these compact binary systems, and then inform the design of 
neural network models~\cite{Adam:2018prd,2020arXiv201203963W,Wei:2020sfz,2022arXiv220911146S}, matched filtering approaches~\cite{2016CQGra..33u5004U,Sachdev:2019vvd}, or 
unmodeled searches~\cite{Sergey:2008CQG,Sergey:2016} to 
discover moderately and highly 
eccentric spinning binaries in future 
discovery campaigns.

We also found that including 
higher order terms will enhance the detectability as the results suggest that the $(\ell=|m|=2)$ modes do not faithfully capture the dynamics of the system for asymmetric mass-ratio systems.

This set of simulations extends the library of open-source simulations introduced in~\cite{huerta_nr_catalog}, stored in the DataVault repository maintained by NCSA at the University of Illinois~\cite{Luo_datavault_2022}. We intend to make this set of simulations publicly available on the same repository soon and until then, any data can be availed upon request to the authors of this paper.

\acknowledgements

We thank the anonymous reviewer for their comments and suggestions,
we thank Rossella Gamba for help in setting up the TEOBResumS model.
This material is based upon work supported by the National Science Foundation under Grants No. NSF-2004879, NSF-1550514, ACI-1238993.
Numerical simulations used compute resources provided by XSEDE under allocation TG-PHY160053. EAH gratefully acknowledges 
support from NSF award OAC-1931561. This material is 
based upon work supported by Laboratory Directed 
Research and Development (LDRD) funding from Argonne 
National Laboratory, provided by the Director, 
Office of Science, of the U.S. Department of Energy 
under Contract No. DE-AC02-06CH11357. 
Plots were produced using matplotlib~\cite{Hunter:2007,thomas_a_caswell_2021_5194481}. Part of the analysis was parallelized using TACC's launcher utility \cite{Wilson:2014:LSF:2616498.2616534}.

\bibliography{references}

\appendix

\section{CONVERGENCE}\label{app:convergence}
\begin{figure}
    \centering
    \includegraphics[width=\linewidth]{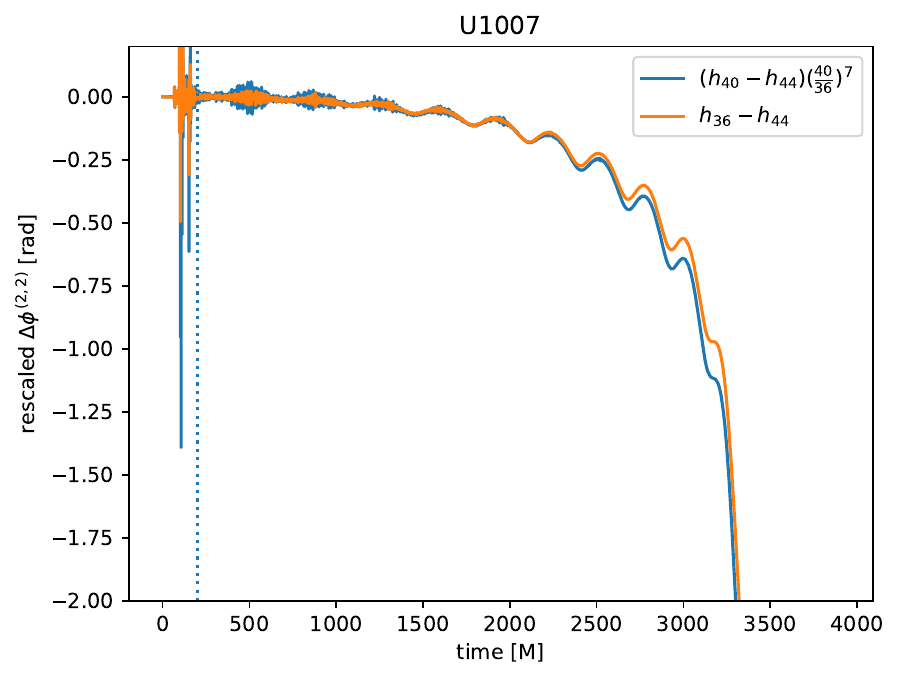}
    \caption{Convergence of the phase difference between the waveform of the highest resolution ($h_{44}$) and the lower resolutions ($h_{40}$, $h_{36}$) with appropriate scaling to get a rough match. This suggests an order of around 7 but this is not representative of the entire library. Note that the plot includes the initial junk radiation (left of vertical dotted line) and the merger and ringdown signal both of which have very large phase differences that are cut out in the plot.}
    \label{fig:convergence_U1007}
\end{figure}

Details about the nature of convergence can be found in appendix B of~\cite{huerta_nr_catalog}. To summarize, though the spatial finite difference operators are at 8th order, the error in the simulations does not scale to 8th order with spatial resolution. This is due to a combination of lower order operations due interpolation on the mesh refinement boundary (5th order accuracy), adaptive mesh refinement operations and varying temporal resolution (from differing spatial resolutions). For each simulation in the library, we have 3 different resolutions which we use to check for convergence\textemdash $N=36,40,44$ corresponding to the number of points in the finest grid radius. We compare the phase difference between the highest resolution and the lower resolutions. To see how much the phase difference reduces with resolution, we scale the phase difference of the higher difference ($h_{40} - h_{44}$) to match the lower difference ($h_{36} - h_{44}$).  Fig.~\ref{fig:convergence_U1007} shows the phase difference of the signals for the U1007 simulation. The order appears to be around 7 which is reasonable for these simulations. Note that this is not the same scaling for other simulations in the library\textemdash it can vary from 4 to 10. This illustrates the point that it is difficult to pull out a universal convergence scaling of the simulations.

\section{INFERRED PARAMETERS FROM NRHybSur2dq8}\label{app:surrogate}

Here we list the inferred parameters from the parameter survey of the NRHybSur2dq8 library of quasicircular, spin-aligned binary waveforms for both the $(\ell=|m|=2)$ and the $l\leq4$ modes separately. The simulations not listed in Table \ref{tab:degeneracy_params} had consistently low FFs across all parameter space. The resolution of the grid search was 0.1 in q, and 0.02 in spins near the inferred values (a lower resolution search was initially done followed by a finer search).

\begin{table}[h]
    \centering
    \begin{tabular}{l|c|c|c|c|c|c|c|c}
         Simulation & \multicolumn{4}{c |}{$\ell=m=2$} & \multicolumn{4}{c}{$l\leq4$} \\
          & q & $\chi_{1z}$ & $\chi_{2z}$ & FF(\%) & q & $\chi_{1z}$ & $\chi_{2z}$ & FF(\%)  \\ \hline
          U0010 & 2.7 & 0.60 & -0.53 & 60.9 & 2.3 & 0.55 & 0.37 & 46.7 \\ \hline
          U0011 & 3.0 & 0.58 & 0.28 & 47.8 & 1.1 & 0.13 & 0.54 & 49.8 \\ \hline
          U0014 & 4.3 & -0.62 & -0.54 & 69.0 & 4.3 & -0.62 & -0.52 & 67.8 \\ \hline
          U0020 & 6.0 & -0.62 & -0.56 & 84.9 & 5.2 & -0.6 & -0.09 & 82.5 \\ \hline
          U0021 & 5.4 & 0.60 & -0.31 & 52.5 & 5.7 & 0.47 & 0.58 & 48.5 \\ \hline
          U0023 & 4.7 & 0.60 & 0.20 & 70.1 & 4.7 & 0.60 & 0.22 & 66.1 \\ \hline
          U1007 & 1.4 & 0.56 & 0.66 & 93.9 & 1.3 & 0.58 & 0.66 & 93.9 \\ \hline
          U1008 & 1.4 & -0.49 & -0.56 & 80.2 & 1.0 & -0.52 & -0.43 & 79.3 \\ \hline
          U1019 & 5.6 & 0.60 & 0.35 & 90.8 & 5.5 & 0.60 & 0.51 & 90.0 \\ \hline
          U0032 & 5.0 & -0.62 & 0.28 & 40.0 & 4.5 & -0.51 & -0.23 & 39.6 \\ \hline
          U0038 & 5.0 & -0.58 & -0.17 & 39.5 & 6.8 & -0.47 & -0.66 & 42.0 \\ \hline
          U0040 & 5.7 & -0.33 & -0.05 & 88.4 & 5.2 & -0.20 & -0.64 & 85.7 \\ \hline
          U0041 & 5.0 & 0.47 & 0.01 & 62.1 & 5.0 & 0.47 & -0.15 & 55.5 \\ \hline
    \end{tabular}
    \caption{Parameters from NRHybSur2dq8 that best match the simulation data along with fitting factors (FFs) for 
    both $(\ell=|m|=2)$ and the $l\leq4$ modes. Low FFs indicate 
    that spin-aligned eccentric signals, such as those in 
    our simulation library, will be poorly recovered or go missing when using current quasicircular template match filtering techniques. The only chance to see these signals would be through unmodeled searches if they are sufficiently loud.}
    \label{tab:degeneracy_params}
\end{table}

\end{document}